\documentclass[preprint,aps]{revtex4}

\usepackage{epsfig}
\def\ffrac#1#2{\textstyle{#1\over#2}\displaystyle}
\begin{document}
\title{Entanglement Entropy and Quantum Field Theory}
\author{Pasquale Calabrese$^1$ and John Cardy$^{1,2}$}
\affiliation{$^1$Rudolf Peierls Centre for Theoretical Physics,
         1 Keble Road, Oxford OX1 3NP, U.K.\\
         $^2$All Souls College, Oxford.}
%
%

%
\begin{abstract}
We carry out a systematic study of entanglement entropy in
relativistic quantum field theory. This is defined as the von
Neumann entropy $S_A=-{\rm Tr}\,\rho_A\log\rho_A$ corresponding to
the reduced density matrix $\rho_A$ of a subsystem $A$. For the
case of a 1+1-dimensional critical system, whose continuum limit
is a conformal field theory with central charge $c$, we re-derive
the result $S_A\sim(c/3)\log\ell$ of Holzhey et al. when $A$ is a
finite interval of length $\ell$ in an infinite system, and extend
it to many other cases: finite systems, finite temperatures, and
when $A$ consists of an arbitrary number of disjoint intervals
({\bf See note added}). For such a system away from its critical
point, when the correlation length $\xi$ is large but finite, we
show that $S_A\sim{\cal A}(c/6)\log\xi$, where $\cal A$ is the
number of boundary points of $A$. These results are verified for a
free massive field theory, which is also used to confirm a scaling
ansatz for the case of finite-size off-critical systems, and for
integrable lattice models, such as the Ising and XXZ models, which
are solvable by corner transfer matrix methods. Finally the
free-field results are extended to higher dimensions, and used to
motivate a scaling form for the singular part of the entanglement
entropy near a quantum phase transition.
\end{abstract}
%

\maketitle


\section{Introduction.}

Recently there has been considerable interest in formulating measures of
quantum entanglement and applying them to extended quantum systems with
many degrees of freedom, such as quantum spin chains.

One of these measures\cite{Bennett}
is entanglement \em entropy\em. Suppose the whole
system is in a pure quantum state $|\Psi\rangle$, with density matrix
$\rho=|\Psi\rangle\langle\Psi|$, and an observer A measures only a
subset $A$ of a complete set of commuting observables, while another
observer B may measure the remainder. A's reduced density matrix is
$\rho_A={\rm Tr}_B\,\rho$. The entanglement entropy is just the von
Neumann entropy $S_A=-{\rm Tr}_A\,\rho_A\log\rho_A$ associated with this
reduced density matrix. It is easy to see that $S_A=S_B$. For an
unentangled product state, $S_A=0$. Conversely, $S_A$ should be a
maximum for a maximally entangled state.

For example, for a system with
two binary (spin-$\frac12$) degrees of freedom, with
$|\Psi\rangle=\cos\theta|\uparrow\downarrow\rangle+\sin\theta|\downarrow
\uparrow\rangle$, where A observes only the first spin and B the second,
$S_A$ takes its maximum value of $\log 2$ when $\cos^2\theta=\frac12$, which
agrees with our intuitive idea of maximal entanglement.
For this system it has been shown,\cite{Bennett} even in the partially
entangled case, that if there are $M$
copies of the state available, by making only local operations A can
produce $M'<M$ states which are maximally entangled. The optimal
conversion ratio $M'/M$ is given, for large $M$, by $S_A$.\footnote{As
far as we are aware, this analysis has not been extended to systems with
many degrees of freedom such as we consider in this paper. Indeed, we
shall see that for such systems $S_A$ can be much larger than unity, so
cannot have such a simple interpretation.}

Although there are other measures of entanglement,\cite{othermeasures}
the entropy is most readily suited to analytic investigation.
In several papers\cite{Vidal,Korepin,leb,Casini},
the concept has been applied to quantum spin chains.
Typically, the subset $A$ consists of a commuting set of components of
the spin degrees of freedom in some interval of length $\ell$, in an
infinitely long chain. It is found that the entanglement entropy
generally tends to a finite value as $\ell$ increases, but that this
value diverges as the system approaches a quantum critical point.
At such a critical point, the entropy grows proportional to $\log\ell$
for large $\ell$.

Close to a quantum critical point, where the correlation length $\xi$ is
much larger than the lattice spacing $a$, the low-lying excitations and
the long-distance behaviour of the correlations in the ground state of
a quantum spin chain are believed to be described by a quantum field
theory in 1+1 dimensions.
If the dispersion relation of the low-lying excitations is linear for
wave numbers $k$ such that $\xi^{-1}\ll|k|\ll a^{-1}$, the field theory
is relativistic. We shall consider only those cases in this paper.
At the critical point, where $\xi^{-1}=0$, the field theory is massless,
and is a \em conformal \em field theory (CFT).

In this case, the von Neumann entropy of subsystem corresponding to an
interval of length $\ell$ was calculated some time ago by Holzhey et
al.\cite{Holzhey},
in the context of black hole physics (although that connection
has been questioned), where it was termed `geometric' entropy. Using
methods of conformal field theory, based in part on earlier work of
Cardy and Peschel\cite{CardyPeschel}, they found
$S_A\sim (c/3)\log(\ell/a)$, where $c$ is the conformal anomaly number
(sometimes called the central charge) of the corresponding CFT.

This result has been verified by analytic and numerical calculations on
integrable quantum spin chains corresponding to CFTs with $c=\frac12$
and $c=1$. \cite{Vidal,Korepin,leb}

In this paper, we first put the CFT arguments of Holzhey et al.\cite{Holzhey}
on a more
systematic basis, and generalise their result in a number of ways.
Our methods are based on a formula for the entropy in terms of the
partition function in the
path integral formulation of the quantum theory as a euclidean
field theory on an $n$-sheeted Riemann surface, in the limit $n\to1$.

For a 1+1-dimensional theory at a critical point, we derive analogous
formulas for the entropy in the cases when the subsystem $A$ consists
of an \em arbitrary \em number of disjoint intervals of the real line (see
Eq.~\ref{general}), and when the whole system has itself a finite length $L$.
For example, for the case when $A$ is a single interval of length
$\ell$, and periodic boundary conditions are imposed on the whole
system, we find
\begin{equation}
S_A=(c/3)\log\big((L/\pi a)\sin(\pi\ell/L)\big)+c_1'\,.
\end{equation}
On the other hand, for a finite system of total length $L$
with open boundaries, divided at some interior point into an interval
$A$ of length $\ell$ and its complement, we find
\begin{equation}
S_A=(c/6)\log\big((L/\pi a)\sin(\pi\ell/L)\big)+2g+c_1'\,,
\end{equation}
where $g$ is the boundary entropy of Affleck and
Ludwig\cite{AffleckLudwig}.
We also treat the case when the system is infinitely
long but is in a thermal mixed state at finite temperature $\beta^{-1}$:
\begin{equation}
\label{finiteT}
S_A=(c/3)\log\big((\beta/\pi a)\sinh(\pi\ell/\beta)\big)+c_1'\,.
\end{equation}
In all these cases, the constant $c_1'$ is the same, but
non-universal.

For a massive 1+1-dimensional relativistic QFT (which corresponds to an
off-critical quantum spin chain where the correlation length $\xi\gg a$)
the simplest results are for an infinite system divided at some point
into two semi-infinite pieces.
In this case we verify that
the entanglement entropy is finite, and derive the universal formula
\begin{equation}
\label{noncrit}
S_A\sim (c/6)\log(\xi/a)\,.
\end{equation}
In the more general case when $A$ consists of a collection of disjoint
intervals, each of length $\gg\xi$, we expect (\ref{noncrit}) to be
multiplied by a factor
$\cal A$ which counts the number of boundary points between $A$ and
$B$ (the 1d analogue of surface area.)

For ${\cal A}=1$ the entropy is exactly
calculable in the case of a free field theory. We verify the above
formula, and exhibit the finite-size
cross-over which occurs when $\xi$ is of the size $L$ of the system.
For a lattice model in this geometry, with $L$ infinite, we point out that
$\rho_A$ is simply related to Baxter's corner transfer matrix, and thus,
for integrable models whose weights satisfy a Yang-Baxter relation, all
its eigenvalues can be determined exactly. We treat explicitly the case
of the Ising model and its anisotropic limit, the transverse Ising spin
chain, and also the XXZ model,
computing exactly at all values of the coupling the finite
part of the entropy $S$. This agrees with our continuum result
(\ref{noncrit}) for $c=\frac12,1 $ when the correlation length $\xi$ is large.

The analysis for the free theory is straightforward to extend
to higher dimensions, at least in suitable geometries. This leads to the
well-known law, first found by Srednicki\cite{s-93}, that the
entropy should be proportional to the surface area $\cal A$ of the
subsystem $A$. As pointed out by Srednicki, the coefficient of this term,
for $d>1$, depends on the UV cut-off and is therefore non-universal.
However, based on our calculations, we propose that there should be a
non-leading piece in $S_A/{\cal A}$, proportional to $\xi^{-(d-1)}$,
which depends
in a singular way on the couplings near a quantum phase transition, and
whose form is, moreover, \em universal\em.

The layout of this paper is as follows. In the next section, we discuss
the entropy in terms of the euclidean path integral on an $n$-sheeted
Riemann surface. In Sec.~\ref{sectcft} we consider the 1+1-dimensional
conformal case. We use the powerful methods of CFT
to show that the partition function on the Riemann surface of interest
is given, up to a constant, by a calculable correlation function of
vertex operators in a $c=1$ CFT in the complex plane. Similar results
apply to a system with boundaries. This general result allows us to
derive all the special cases described above. In Sec.~\ref{sectmass} we
consider the case of a massive 1+1-dimensional field theory. We derive
the result $S\sim(c/6)\log(\xi/a)$ from completely general properties of
the stress tensor in the relevant geometry. This is then verified for a
free bosonic massive field.
In the last part of this section we relate the
lattice version of this problem to the corner transfer matrix, and
compute the off-critical entropy for the case of the Ising and
XXZ Heisenberg spin chains.
In Sec.~\ref{sectfss} we study the off-critical case in a finite-size
system, propose a scaling law, and verify it for the case of a free
massive field theory. We compute the relevant scaling function in a
systematic sequence of approximations.
Sec.~\ref{secthd} is devoted to the discussion of higher dimensions.

While all this work was being carried out, some other related papers
have appeared in the literature. In Ref.~\cite{Korepin2}, the
result $S_A\sim(c/3)\log(\ell/a)$, first found by Holzhey et
al.\cite{Holzhey},
was obtained by the following argument: it was assumed
that the entropy should be conformally invariant and therefore some
function $F(x)$ of the variable $x=(\beta/\pi)\sinh(\pi\ell/\beta)$;
by comparing with known case $\ell\gg\beta$ \cite{BCN,Affleck} it was
observed that $F(x)$ should behave as $(c/3)\log x$ as $x\to\infty$;
finally it was assumed (with no justification being given) that the
$\log x$ form for $F$ is valid for all values of $x$; finally
the limit $\beta\gg\ell$ was taken. In the present paper, we should
stress, we have derived all these statements from first principles of
CFT.

Very recently Casini and Huerta\cite{Casini} have considered the case
of two intervals $A$ and $B$, and argued that the quantity
$F(A,B)\equiv S(A)+S(B)-S(A\cap B)-S(A\cup B)$ is UV finite as $a\to0$,
and is given by a universal logarithmic function of the cross-ratio
of the four end points. This corresponds to, and agrees with, our case $N=2$.
These authors, however, assume the conformal invariance of the entropy,
while, once again, we stress that in the present paper we derive this from
fundamental
properties of the stress tensor. These authors also give a very nice
alternative derivation of Zamolodchikov's $c$-theorem\cite{Zam} based on
this quantity $F(A,B)$.

\section{von Neumann Entropy and Riemann surfaces.}
\label{sectrs}
Consider a lattice quantum theory in one space and one time
dimension, initially on the infinite line. The lattice spacing is $a$,
and the lattice sites are labelled by a discrete variable $x$.
The domain of $x$ can be finite, i.e. some interval of length $L$,
semi-infinite, or infinite. Time is considered to be continuous.
A complete set of local
commuting observables will be denoted by $\{\hat\phi(x)\}$, and their
eigenvalues and corresponding eigenstates by $\{\phi(x)\}$ and
$\otimes_x|\{\phi(x)\}\rangle$ respectively. For a bosonic lattice
field theory,
these will be the fundamental bosonic fields of the theory; for a spin
model some particular component of the local spin. The dynamics of the
theory is described by a time-evolution operator $\hat H$. The density
matrix $\rho$ in a thermal state at inverse temperature $\beta$
is
\begin{equation}
\rho(\{\phi(x'')''\}|\{\phi(x')'\})=
Z(\beta)^{-1}\langle\{\phi(x'')''\}|e^{-\beta\hat
H}|\{\phi(x')'\}\rangle\,,
\end{equation}
where $Z(\beta)={\rm Tr}\,e^{-\beta\hat H}$ is the partition function.

This may be expressed in the standard way
as a (euclidean) path integral:
\begin{equation}
\label{pathi}
\rho=Z^{-1}\int[d\phi(x,\tau)]
\prod_x\delta(\phi(x,0)-\phi(x')')\prod_x
\delta(\phi(x,\beta)-\phi(x'')'')\,e^{-S_E}\,,
\end{equation}
where $S_E=\int_0^\beta L_Ed\tau$, with $L_E$ the euclidean lagrangian.
(For a spin model this would be replaced by a coherent state path
integral.)

The normalisation factor of the partition function ensures that
${\rm Tr}\,\rho=1$, and is found by
setting $\{\phi(x)''\}=\{\phi(x)'\}$ and integrating
over these variables. This has the effect of sewing together the edges
along $\tau=0$ and $\tau=\beta$ to form a cylinder of circumference
$\beta$.

Now let $A$ be a subsystem consisting of the points $x$ in the
disjoint\footnote{This restriction is not necessary, but the set-up is
easier to picture in this case.}
intervals $(u_1,v_1),\ldots,(u_N,v_N)$. An expression for the
the reduced density matrix $\rho_A$
may be found from (\ref{pathi})
by sewing together only those points $x$ which are not in $A$. This
will have the effect of leaving open cuts, one for each interval
$(u_j,v_j)$, along the the line $\tau=0$.

We may then compute ${\rm Tr}\,\rho_A^n$, for any positive
integer $n$, by making $n$ copies of the above, labelled by an integer
$k$ with $1\leq k\leq n$, and sewing them together cyclically along the
the cuts so that $\phi(x)'_k=\phi(x)''_{k+1}$ (and
$\phi(x)'_n=\phi(x)''_1$) for all $x\in A$. Let us denote the
path integral on this $n$-sheeted structure by $Z_n(A)$.
Then
\begin{equation}
\label{ZoverZ}
{\rm Tr}\,\rho_A^n={Z_n(A)\over Z^n}\,.
\end{equation}

Now, since ${\rm Tr}\,\rho_A^n=\sum_\lambda\lambda^n$,
where $\lambda$ are the eigenvalues of $\rho_A$ (which lie in
$[0,1)$,) and
since ${\rm Tr}\,\rho_A=1$, it follows that
the left hand side is absolutely convergent and therefore
analytic for all ${\rm Re}\,n>1$. The derivative wrt $n$ therefore also
exists and is analytic in the region. Moreover, if the entropy
$\rho_A=-\sum_\lambda\lambda\log\lambda$ is finite, the limit as
$n\to1+$ of the first derivative converges to this value.

We conclude that the right hand side of (\ref{ZoverZ}) has a unique
analytic continuation to ${\rm Re}\,n>1$ and that its first derivative
at $n=1$ gives the required entropy:
\begin{equation}
S_A=-\lim_{n\to1}{\partial\over\partial n}{Z_n(A)\over
Z^n}\,.
\end{equation}
(Note that even before taking this limit, (\ref{ZoverZ}) gives an
expression for the Tsallis entropy\cite{Tsallis}
$({\rm Tr}\,\rho_{A}^n-1)/(1-n)$.)

So far, everything has been in the discrete space domain. We now discuss
the continuum limit, in which $a\to0$ keeping all other lengths fixed.
The points $x$ then assume real values, and the path integral is over
fields $\phi(x,\tau)$ on an $n$-sheeted Riemann surface, with branch
points at $u_j$ and $v_j$.
In this limit, $S_E$ is supposed to go over into the euclidean action for
a quantum field theory. We shall restrict attention to the case when
this is Lorentz invariant, since the full power of relativistic field
theory can then be brought to bear.
The behaviour of partition functions in this limit has been well
studied. In two dimensions, the logarithm of a general partition function
$Z$ in a domain with total area $\cal A$ and with boundaries of total
length $\cal L$ behaves as
\begin{equation}
\label{divs}
\log Z=f_1{\cal A}a^{-2}+f_2{\cal L}a^{-1}+\ldots
\end{equation}
where $f_1$ and $f_2$ are the non-universal bulk and boundary free
energies. Note, however, that these leading terms \em cancel \em in the
ratio of partition functions in (\ref{ZoverZ}).
However, as was argued by Cardy and Peschel\cite{CardyPeschel}
there are also \em universal \em terms proportional to $\log a$. These
arise from points of non-zero curvature of the manifold and its
boundary. In our case, these are conical singularities at the branch
points. In fact, as we shall show, it is precisely these logarithmic
terms which give rise to the non-trivial dependence of the final result
for the entropy on the short-distance cut-off $a$. For the moment let
us simply remark that, in order to achieve a finite limit as $a\to0$,
the right hand side of (\ref{ZoverZ}) should be multiplied by some
renormalisation constant ${\cal Z}(A,n)$. Its dependence on $a$
will emerge from the later analysis.

\def\be{\begin{equation}}
\def\ee{\end{equation}}
\section{Entanglement entropy in 2d conformal field theory.}
\label{sectcft}

Now specialise the discussion of the previous section
to the case when the field theory is relativistic and massless, i.e. a
conformal field theory (CFT), with central charge $c$, and initially consider
the case of zero temperature.

We show that in this case the ratio of partition functions in
(\ref{ZoverZ}) is the same as the correlation function arising from the
insertion of primary
scaling operators $\Phi_n(u_j)$ and $\Phi_{-n}(v_j)$, with
scaling dimensions $X_n=2\Delta_n=(c/12)(1-1/n^2)$, into each of the $n$
(disconnected) sheets. Moreover, this $2N$-point correlation function
is computable from the Ward identities of CFT.

In the language of string theory, the objects we consider are
correlators of orbifold points in theories whose target
space consists of $n$ copies of the given CFT. We expect that some of
our results may therefore have appeared in the literature of the subject.

\subsection{Single interval}
We first consider the case $N=1$ and no boundaries, that is the case
considered by Holzhey et al.\cite{Holzhey} of
a single interval of length $\ell$ in an infinitely long 1d quantum
system, at zero temperature.
The conformal mapping
$w\to\zeta=(w-u)/(w-v)$ maps the branch points to $(0,\infty)$. This is
then uniformised by the mapping $\zeta\to z=\zeta^{1/n}=
\big((w-u)/(w-v)\big)^{1/n}$. This maps the whole of the $n$-sheeted
Riemann surface ${\cal R}_n$ to the $z$-plane $\bf C$.
Now consider the holomorphic component
of the stress tensor $T(w)$. This is related to the transformed stress
tensor $T(z)$ by\cite{BPZ}
\begin{equation}
\label{schwartz}
T(w)=(dz/dw)^2\,T(z)+\ffrac c{12}\{z,w\}\,,
\end{equation}
where $\{z,w\}$ is the Schwartzian derivative
$(z'''z'-\frac32{z''}^2)/{z'}^2$.
In particular, taking the expectation value of this, and using $\langle
T(z)\rangle_{\bf C}=0$ by translational and rotational invariance, we find
\begin{equation}
\langle T(w)\rangle_{{\cal R}_n}=\frac c{12}\{z,w\}=
{c(1-(1/n)^2)\over 24}{(v-u)^2\over(w-u)^2(w-v)^2}\,.
\end{equation}
This is to be compared with the standard form\cite{BPZ} of the correlator of
$T$ with two primary operators $\Phi_n(u)$ and $\Phi_{-n}(v)$ which have
the same complex scaling dimensions $\Delta_n=\overline\Delta_n=
(c/24)(1-(1/n)^2)$:
\begin{equation}
\label{TPP}
\langle T(w)\Phi_n(u)\Phi_{-n}(v)\rangle_{\bf C}=
{\Delta_n\over (w-u)^2(w-v)^2(v-u)^{2\Delta_n-2}
(\bar v-\bar u)^{2\overline\Delta_n}}\,,
\end{equation}
where $\Phi_{\pm n}$ are normalised so that
$\langle \Phi_n(u)\Phi_{-n}(v)\rangle_{\bf C}=|v-u|^{-2\Delta_n
-2\overline\Delta_n}$. (\ref{TPP}) is equivalent to the
conformal Ward identity\cite{BPZ}
\begin{equation}
\label{WI}
\langle T(w)\Phi_n(u)\Phi_{-n}(v)\rangle_{\bf C}=
\left({\Delta_n\over(w-u)^2}+{\Delta_n\over(w-v)^2}
+{1\over w-u}{\partial\over\partial u}+
{1\over w-v}{\partial\over\partial v}\right)
\langle\Phi_n(u)\Phi_{-n}(v)\rangle_{\bf C}\,.
\end{equation}
In writing the above, we are assuming that $w$ is a complex coordinate
on a single sheet $\bf C$, which is now decoupled from the others.
We have therefore shown that
\begin{equation}
\langle T(w)\rangle_{{\cal R}_n}\equiv
{\int[d\phi]T(w)e^{-S_E({\cal R}_n)}\over
\int[d\phi]e^{-S_E({\cal R}_n)}}=
{\langle T(w)\Phi_n(u)\Phi_{-n}(v)\rangle_{\bf C}
\over \langle \Phi_n(u)\Phi_{-n}(v)\rangle_{\bf C}}\,.
\end{equation}
Now consider the effect of an infinitesimal conformal transformation
$w\to w'=w+\alpha(w)$ of $\bf C$ which
act identically on all the sheets of ${\cal R}_n$.
The effect of this is to insert a factor
\begin{equation}
{1\over 2\pi i}\int_C\alpha(w)T(w)dw
-{1\over 2\pi i}\int_C\overline{\alpha(w)}\overline T(\bar w)d\bar w
\end{equation}
into the path integral, where the contour $C$ encircles the points $u$
and $v$. The insertion of $T(w)$ is given by (\ref{TPP}). Since this is
to be inserted on each sheet, the right hand side gets multiplied by a
factor $n$.

Since the Ward identity (\ref{WI})
determines all the properties under conformal transformations, we
conclude that the renormalised
$Z_n({A})/Z^n\propto{\rm Tr}\,\rho_{A}^n$ behaves
(apart from a possible overall constant)
under scale and conformal transformations identically to the $n$th power
of two-point function
of a primary operator $\Phi_n$ with $\Delta_n=\overline\Delta_n=
(c/24)(1-(1/n)^2)$. In particular, this means that
\begin{equation}
{\rm Tr}\,\rho_{A}^n=c_n\big((v-u)/a)^{-(c/6)(n-1/n)}\,,
\end{equation}
where the exponent is just $4n\Delta_n$. The power of $a$
(corresponding to the renormalisation constant $\cal Z$) has been
inserted so as the make the final result dimensionless, as it should
be. The constants $c_n$ are not determined by this method. However $c_1$
must be unity. Differentiating with respect to $n$ and setting $n=1$, we
recover the result of Holzhey et al.

The fact that ${\rm Tr}\,\rho_{A}^n$ transforms under a general
conformal transformation as a 2-point function of primary operators
$\Phi_{\pm n}$
means that it is simple to compute in other geometries, obtained by a
conformal mapping $z\to z'=w(z)$,
using the formula\cite{BPZ}
\be
\langle\Phi(z_1,\bar z_1)\Phi(z_2,\bar z_2)\ldots\rangle
=\prod_j|w'(z_j)|^{2\Delta_n}
\langle\Phi(w_1,\bar w_1)\Phi(w_2,\bar w_2)\ldots\rangle\,.
\ee

For example,
the transformation $w\to w'=(\beta/2\pi)\log w$ maps each sheet in
the $w$-plane into an infinitely long cylinder of circumference $\beta$.
The sheets are now sewn together along a branch cut joining the images
of the points $u$ and $v$. By arranging this to lie parallel to the axis
of the cylinder, we get an expression for ${\rm Tr}\,\rho_{A}^n$
in a thermal mixed state at finite temperature $\beta^{-1}$.
This leads to the result for the entropy
\begin{equation}
S_A(\beta)\sim(c/3)\log\big((\beta/\pi a)\sinh(\pi\ell/\beta)\big)+c_1'\,.
\end{equation}
For $\ell\ll\beta$ we find $S_A\sim(c/3)\log(\ell/a)$ as before, while, in the
opposite limit $\ell\gg\beta$, $S_A\sim(\pi c/3)(\ell/\beta)$. In this
limit, the von Neumann entropy is extensive, and its density agrees with
that of the Gibbs entropy of an isolated system of length $\ell$,
as obtained from the standard CFT expression\cite{BCN,Affleck}
$\beta F\sim -(\pi c/6)(\ell/\beta)$ for its free energy.

Alternatively, we may orient the branch cut perpendicular to the axis
of the cylinder, which, with the replacement $\beta\to L$, corresponds
to the entropy of a subsystem of length $\ell$ in a finite 1d system of
length $L$, with periodic boundary conditions, in its ground state.
This gives
\begin{equation}
S_A\sim (c/3)\log\big((L/\pi a)\sin(\pi\ell/L)\big)+c_1'\,.
\end{equation}
Note that this expression is symmetric under $\ell\to L-\ell$. It is
maximal when $\ell=L/2$.

\subsection{Finite system with a boundary.}
Next consider the case when the 1d system is a semi-infinite line,
say $[0,\infty)$, and the subsystem $A$ is the finite interval
$[0,\ell)$. The $n$-sheeted Riemann surface then consists of $n$ copies
of the half-plane $x\geq0$, sewn together along $0\leq x<\ell, \tau=0$.
Once again, we work initially at zero temperature. It is convenient
to use the complex variable $w=\tau+ix$. The uniformising transformation
is now $z=\big((w-i\ell)/(w+i\ell)\big)^{1/n}$, which maps the whole
Riemann surface to the unit disc $|z|\leq1$. In this geometry,
$\langle T(z)\rangle=0$ by rotational invariance, so that, using
(\ref{schwartz}), we find
\begin{equation}
\label{hp}
\langle T(w)\rangle_{{\cal R}_n}=
{\Delta_n(2\ell)^2\over(w-i\ell)^2(w+i\ell)^2}\,,
\end{equation}
where $\Delta_n=(c/24)(1-n^{-2})$ as before.
Note that in the half-plane, $T$ and $\overline T$ are related by
analytic continuation: $\overline T(\bar w)=T(w)^*$.
(\ref{hp}) has the same form as $\langle T(w)\Phi_n(i\ell)\rangle$,
which follows from the Ward identities of boundary CFT\cite{JCbound}, with the
normalisation $\langle\Phi_n(i\ell)\rangle=(2\ell)^{-\Delta_n}$.

The analysis then proceeds in analogy with the previous case. We find
\begin{equation}
{\rm Tr}\,\rho_{A}^n\sim \tilde c_n(2\ell/a)^{(c/12)(n-1/n)}\,,
\end{equation}
so that
$S_A\sim(c/6)\log(2\ell/a)+{\tilde c}'_1$.

Once again, this result can be conformally transformed into a number of
other cases. At finite temperature $\beta^{-1}$ we find
\begin{equation}
S_A(\beta)\sim(c/6)\log\big((\beta/\pi a)\sinh(2\pi\ell/\beta)\big)+
{\tilde c}'_1\,.
\end{equation}
By taking the limit when $\ell\gg\beta$ we find the same extensive
entropy as before. However, we can now identify ${\tilde c}'_1-c'_1$
as the boundary entropy $g$, first discussed by
Affleck and Ludwig\cite{AffleckLudwig}.

For a completely finite 1d system, of length $L$, at
zero temperature, divided into two pieces of lengths $\ell$ and
$L-\ell$, we similarly find
\begin{equation}
S_A=(c/6)\log\big((L/\pi a)\sin(\pi\ell/L)\big)+2g+c'_1\,.
\end{equation}

\subsection{General case.}

For general $N$, the algebra is more complicated, but the method is the
same. The uniformising transformation now has the form
$z=\prod_i(w-w_i)^{\alpha_i}$, with
$\sum_i\alpha_i=0$ (so there is no singularity at infinity.) Here
$w_i$ can be $u_j$, $v_j$, or $\pm iu_j$ in the case of a boundary.
In our case, we have $\alpha_i=\pm1/n$, but it is interesting to
consider the more general transformation, and the notation is simpler.
Once again we have
\be
\langle T(w)\rangle=\frac c{12}\{z,w\}=
\frac c{12}{z'''z'-\ffrac32{z''}^2\over {z'}^2}\,.
\ee
Consider $\{z,w\}/z$.
As a function of $w$, this is meromorphic, has a double pole at each $w=w_i$,
and is $O(w^{-2})$ as $w\to\infty$. Hence it has the form
\be
\{z,w\}/z=\sum_i{A_i\over(w-w_i)^2}+\sum_i{B_i\over w-w_i}\,,
\ee
where $\sum_iB_i=0$. In order to determine $A_i$ and $B_i$, we need
to compute $z'$, etc, to second order in their singularities at $w_i$.
After some algebra, we find
\begin{eqnarray}
z'&=&
\left[{\alpha_i\over w-w_i}+\sum_{j\not=i}{\alpha_j\over w-w_j}\right]\,z\,;\\
z''&=&\left[{-\alpha_i+\alpha_i^2\over(w-w_i)^2}
+2{\alpha_i\over w-w_i}\sum_{j\not=i}{\alpha_j\over
w_i-w_j}+\cdots\right]\,z\,;\\
z'''&=&\left[{2\alpha_i-3\alpha_i^2+\alpha_i^3\over(w-w_i)^3}
+{-3\alpha_i+3\alpha_i^2\over(w-w_i)^2}\sum_{j\not=i}{\alpha_j\over w_i-w_j}
+\cdots\right]\,z\,.
\end{eqnarray}
Let
\be
C_i\equiv \sum_{j\not=i}{\alpha_j\over w_i-w_j}\,.
\ee
Then the coefficient of $(w-w_i)^{-2}$ is
\begin{eqnarray*}
&&\left\{\left[\alpha_i(\alpha_i-1)(\alpha_i-2)+3\alpha_i(\alpha_i-1)C_i(w-w_i)
+\cdots\right]\left[\alpha_i+C_i(w-w_i)+\cdots\right]\right.\\
&&\left.-\ffrac32\left[\alpha_i(\alpha_i-1)+2\alpha_iC_i(w-w_i)
+\cdots\right]^2\right\}/
\left[\alpha_i+C_i(w-w_i)+\cdots\right]^2\,,
\end{eqnarray*}
from which we find after a little more algebra that
\begin{eqnarray}
A_i &=&\ffrac12(1-\alpha_i^2)\,;\\
B_i &=&C_i\,{1-\alpha_i^2\over \alpha_i}\,.
\end{eqnarray}
Thus we have shown that
\be
\langle T(w)\rangle=\frac c{12}
\sum_i\left[{\ffrac12(1-\alpha_i^2)\over(w-w_i)^2}
+\left({(1-\alpha_i^2)\over\alpha_i}\sum_{j\not=i}{\alpha_j\over w_i-w_j}\right)
{1\over w-w_i}\right]\,.
\ee
This is to be compared with the conformal Ward identity
\be
\langle T(w)\prod_i\Phi_i(w_i)\rangle
=\sum_i\left[{\Delta_i\over(w-w_i)^2}
+{1\over w-w_i}{\partial\over\partial w_i}\right]
\langle\prod_k\Phi_k(w_k)\rangle\,.
\ee
For these to be equivalent, we must have $\Delta_i=\frac12(1-\alpha_i^2)$
and
\be
{(1-\alpha_i^2)\over\alpha_i}\sum_{j\not=i}{\alpha_j\over w_i-w_j}
={\partial \over\partial w_i}\log\langle\prod_k\Phi_k(w_k)\rangle\,.
\ee
A necessary and sufficient condition for this is that
\be
{\partial\over\partial w_k}\left({1-\alpha_i^2\over\alpha_i}
\sum_{j\not=i}{\alpha_j\over w_i-w_j}\right)=
{\partial\over\partial w_i}\left({1-\alpha_k^2\over\alpha_k}
\sum_{j\not=k}{\alpha_j\over w_k-w_j}\right)\,,
\ee
for each pair $(i,k)$. This reduces to
\be
-{1-\alpha_i^2\over \alpha_i}{\alpha_k\over(w_k-w_i)^2}=
-{1-\alpha_k^2\over \alpha_k}{\alpha_i\over(w_i-w_k)^2}\,,
\ee
that is,
$\alpha_i=\pm\alpha_k$
for each pair $(i,k)$. Since $\sum_i\alpha_i=0$ the only way to satisfy
this is to have $\alpha_i=\alpha\sigma_i$, with $\sigma_i=\pm1$,
and half the $\sigma_i=+1$ and the remainder $-1$.
Interestingly enough, this is precisely the case we need, with
$\alpha=1/n$.

If these conditions are satisfied,
\be
{\partial\over\partial w_i}\log\langle\prod_j\Phi_j(w_j)\rangle
=\frac c{12}(1-\alpha^2)\sum_{k\not=i}{\sigma_i\sigma_k
\over w_i-w_k}\,,
\ee
so that
\be
\log\langle\prod_j\Phi_j(w_j)\rangle
=\ffrac c{12}(1-\alpha^2)\sum_{k<i}\sigma_i\sigma_k
\log(w_i-w_k)+ E\,,
\ee
where $E$ is independent of all the $w_i$.
In the case with no boundary, $E$ can depend however on the
$\bar w_i$. A similar calculation with $\overline T$ then gives a
similar dependence. We conclude that ${\rm Tr}\,\rho_{A}^n$
behaves under conformal transformations in the same way as
\be
\langle\prod_j\Phi_j\rangle
\propto\prod_{j<k}
(w_i-w_k)^{\frac c{12}(1-\alpha^2)\sigma_i\sigma_k}
(\bar w_i-\bar w_k)^{\frac c{12}(1-\alpha^2)\bar\sigma_i\bar\sigma_k}\,.
\ee
Taking now $\alpha=1/n$, and $\sigma=\pm1$ according as $w_i=u_j$ or
$v_j$, we find
\begin{equation}
\label{generaln}
{\rm Tr}\,\rho_{A}^n\sim
c_n^N\left({\prod_{j<k}(u_k-u_j)(v_k-v_j)\over\prod_{j\leq k}(v_k-u_j)}
\right)^{(c/6)(n-1/n)}\,.
\end{equation}
The overall constant is fixed in terms of the previously defined $c_n$
by taking the intervals to be far apart
from each other, in comparison to their lengths.

Differentiating with respect to $n$ and setting $n=1$,
we find our main result of this section
\begin{equation}
\label{general}
S_A=\frac c3\left(\sum_{j\leq k}\log\big((v_k-u_j)/a\big)
-\sum_{j<k}\log\big((u_k-u_j)/a\big)-\sum_{j<k}\log\big((v_k-v_j)/a\big)
\right)+Nc_1'\,.
\end{equation}

A similar expression holds in the case of a boundary, with half of the
$w_i$ corresponding to the image points.

Finally we comment on the recent result of Casini and
Huerta\cite{Casini}, which corresponds to $N=2$. In fact, it may be
generalised to the ratio of Tsallis entropies: from (\ref{generaln})
we find
\begin{equation}
{S_n(A)S_n(B)\over S_n(A\cap B)S_n(A\cup B)}=
\left({(v_1-u_1)(v_2-u_2)\over (u_2-u_1)(v_2-v_1)}\right)^{(c/6)(n-1/n)}\,,
\end{equation}
where $S_n(A)\equiv{\rm Tr}\,\rho_A^n$, and
the expression in parentheses is the cross-ratio $\eta$ of the four
points. Notice that in this expression the dependence on
the ultraviolet cut-off $a$ disappeared, so have the non-universal
numbers $c_n$. Differentiating with respect to $n$ gives the result of
Casini and Huerta\cite{Casini}, who however \em assumed \em
that the result should depend only on $\eta$.

\section{Entropy in non-critical 1+1-dimensional models}
\label{sectmass}
\subsection{Massive field theory - general case}
In this section we consider an infinite non-critical
model in 1+1-dimensions, in the scaling limit where the lattice spacing
$a\to0$ with the correlation length (inverse mass) fixed.
This corresponds to a massive relativistic QFT. We first consider the
case when the subset $A$ is the negative real axis, so that the
appropriate Riemann surface has branch points of order $n$ at 0 and
infinity. However, for the non-critical case, the branch point at
infinity is unimportant: we should arrive at the same expression
by considering a finite system whose length $L$ is much greater than
$\xi$.

Our argument parallels that of Zamolodchikov\cite{Zam} for the proof of his
famous $c$-theorem.
Let us consider the expectation value of the stress tensor $T_{\mu\nu}$
of a massive euclidean QFT on such a Riemann surface. In complex
coordinates, there are three non-zero components: $T\equiv T_{zz}$,
$\overline T\equiv T_{\bar z\bar z}$, and the trace
$\Theta=4T_{z\bar z}=4T_{\bar zz}$. These are related by the
conservation equations
\begin{eqnarray}
\label{cons}
\partial_{\bar z}T+\ffrac14\partial_z\Theta&=&0\,;\\
\partial_z\overline T+\ffrac14\partial_{\bar z}\Theta&=&0\,.
\end{eqnarray}

Consider the expectation values of these components. In the
single-sheeted geometry, $\langle T\rangle$ and $\langle\overline
T\rangle$ both vanish, but $\langle\Theta\rangle$ is constant and
non-vanishing: it measures the explicit breaking of scale invariance in
the non-critical system. In the $n$-sheeted geometry, however, they
all acquire a non-trivial spatial dependence. By
rotational invariance about the origin, they have the form
\begin{eqnarray}
\langle T(z,\bar z)\rangle&=& F_n(z\bar z)/z^2\,;\\
\langle\Theta(z,\bar z)\rangle-\langle\Theta\rangle_1&=&
G_n(z\bar z)/(z\bar z)\,;\\
\langle\overline T(z,\bar z)\rangle&=& F_n(z\bar z)/{\bar z}^2 \,.
\end{eqnarray}
From the conservation conditions (\ref{cons}) we have
\begin{equation}
(z\bar z)\left(F'_n+\ffrac14G'_n\right)=\ffrac14G_n\,.
\end{equation}
Now we expect that $F_n$ and $G_n$ both approach zero exponentially fast for
$|z|\gg\xi$, while in the opposite limit, on distance scales $\ll\xi$,
they approach the CFT values (see previous section)
$F_n\to (c/24)(1-n^{-2})$, $G_n\to0$.

This means that if we define an effective $C$-function
\begin{equation}
C_n(R^2)\equiv \left(F(R^2)+\ffrac14G(R^2)\right)\,,
\end{equation}
then
\begin{equation}
R^2{\partial\over\partial(R^2)}C_n(R^2)=\ffrac14G_n(R^2)\,.
\end{equation}

If we were able to argue that $G_n\leq0$, that is
$\langle\Theta\rangle_n\leq\langle\Theta\rangle_1$, we would have
found alternative formulation of the $c$-theorem.
However, we can still derive an integrated form of the $c$-theorem,
using the boundary conditions:\footnote{We have assumed that theory is
trivial in the infrared. If the RG flow is towards a non-trivial theory, $c$
should be replaced by $c_{UV}-c_{IR}$.}
\begin{equation}
\int_0^\infty {G_n(R^2)\over R^2}d(R^2)=-(c/6)(1-n^{-2})\,,
\end{equation}
or equivalently
\begin{equation}
\int\left(\langle\Theta\rangle_n-\langle\Theta\rangle_1\right)d^2\!R
=-\pi n(c/6)(1-n^{-2})\,,
\end{equation}
where the integral is over the whole of the the $n$-sheeted surface.
Now this integral (multiplied by a factor $1/2\pi$ corresponding to the
conventional normalisation of the stress tensor)
measures the response of the free energy $-\log Z$
to a scale transformation, i.e. to a change in the mass $m$, since this is the
only dimensionful parameter of the renormalised theory.
Thus the left hand side is equal to
\begin{equation}
-(2\pi)\,m(\partial/\partial m)\left[\log Z_n-n\log Z\right]\,,
\end{equation}
giving finally
\begin{equation}
{Z_n\over Z^n}= c_n(ma)^{(c/12)(n-1/n)}\,,
\end{equation}
where $c_n$ is a constant (with however $c_1=1$),
and we have inserted a power of $a$,
corresponding to the renormalisation constant $\cal Z$ discussed
earlier, to make the result dimensionless.

This shows that the $(n-1/n)$ dependence for the exponent of the
Tsallis entropy is a
general property of the continuum theory. Differentiating at $n=1$, we
find the main result of this section
\begin{equation}
\label{mass}
S_{A}\sim-(c/6)\log(ma)=(c/6)\log(\xi/a)\,,
\end{equation}
where $\xi$ is the correlation length. We re-emphasise that this result
was obtained only for the scaling limit $\xi\gg a$. However, for lattice
integrable models, we shall show how it is possible to obtain the full
dependence without this restriction.

So far we have considered the simplest geometry in the which set $A$ and
its complement $B$ are semi-infinite intervals. The more general case,
when $A$ is a union of disjoint intervals, is more difficult in the
massive case. However it is still true that the entropy can be expressed
in terms of the derivative at $n=1$ of correlators of operators
$\Phi_n$. The above calculation can be thought of in terms of the
one-point function $\langle\Phi_n\rangle$. In any quantum field theory
a more general correlator
$\langle\prod_{i=1}^{k}\Phi_{\pm n}(w_i)\rangle$ should obey cluster
decomposition: that is, for separations $|w_i-w_j|$ all $\gg\xi$, it
should approach $\langle\Phi_n\rangle^{k}$. This suggests that, in this
limit, the entropy should behave as $S_A\sim{\cal A}(c/6)\log(\xi/a)$,
where ${\cal A}=k$ is the number of boundary points between $A$ and its
complement. This would be the 1d version of the area
law\cite{s-93}. When the interval lengths are of the order of
$\xi$, we expect to see a complicated but universal scaling form for the
cross-over.

\def\be{\begin{equation}}
\def\ee{\end{equation}}

\def\bea{\begin{eqnarray}}
\def\eea{\end{eqnarray}}

\def\e{\epsilon}

\subsection{Free bosonic field theory}

In this subsection we verify Eq.~(\ref{mass}) by an explicit calculation for a
massive free field theory (Gaussian model.)
The action
\be
{\cal S}=\int \ffrac12
\left((\partial_\mu\varphi)^2+m^2\varphi^2\right)d^2\!r,
\ee
is, as before,
considered on a $n$-sheeted Riemann surface with one cut, which we
arbitrarily fix on the real negative axis.

To obtain the entanglement entropy we should know the ratio
$Z_n/Z^n$, where $Z_n$ is the partition function
in the $n$-sheeted geometry. There are several equivalent ways to calculate
such partition function. In the following, we find easier to use the
identity\footnote{This holds only for non-interacting theories: in the
presence of interactions the sum of all the zero-point diagrams has to be
taken into account.}
\be
\frac{\partial}{\partial m^2}\log Z_n=-\ffrac{1}{2}\int
\,G_n({\bf r},{\bf r})d^2\!r\,,
\label{logZ}
\ee
where $G_n({\bf r,r'})$ is the two-point correlation function in the
$n$-sheeted geometry. Thus we need the combination $G_n-nG_1$. $G_n$ obeys
\be
(-\nabla^2_{\bf r}+m^2) G_n({\bf r,r'}) = \delta^2({\bf r-r'}) \,.
\label{Hde}
\ee
Its solution (see the Appendix) may be expressed in polar coordinates as
(here $0<r,r'<\infty$ and $0\leq\theta,\theta'<2\pi n$)
\be
G_n(r,\theta,r',\theta')=
\frac{1}{2\pi n} \sum_{k=0}^\infty d_k \int_0^\infty \lambda d\lambda
\frac{J_{k/n}(\lambda r)J_{k/n}(\lambda r')}{\lambda^2+m^2}
{\cal C}_k(\theta,\theta')\,,
\label{2pt}
\ee
where
${\cal C}_k(\theta,\theta')=
\cos(k\theta/n)\cos(k\theta'/n)+\sin(k\theta/n)\sin(k\theta'/n)$,
$d_0=1$, $d_{k>0}=2$, and $J_k(x)$ are the Bessel functions of the first kind.
At coincident points (i.e. $r=r'$, $\theta=\theta'$),
and after integrating over $\theta$ and $\lambda$ we have
\be
G_n(r)\equiv G_n({\bf r},{\bf r})=\sum_{k=0}^\infty d_k I_{k/n} (m r) K_{k/n} (mr)\,,
\label{Gr}
\ee
where $I_k(x)$ and $K_k(x)$ are the modified Bessel functions of the first
and second kind respectively \cite{as}.

The sum over $k$ in (\ref{Gr}) is UV divergent. This reflects the usual
short-distance divergence which would occur even in the plane. However,
if we formally
exchange the order of the sum and integration we find
\be
-\frac{\partial}{\partial m^2}\log Z_n=
\frac{1}{2}\int d^2 r \,G_n(r) {=}
\frac{1}{2}\sum_k d_k \int_0^\infty I_{k/n} (m r) K_{k/n} (mr) r dr=
\frac{1}{4nm^2}\sum_k d_k k\,.
\label{div}
\ee
Interpreting the last sum as
$2\zeta(-1)=-1/6$,
we obtain the correct result, which we now derive more systematically.

Let us first regularise each sum over $k$ by inserting a function
$F(k/\Lambda_n)$: $F$ is chosen so that $F(0)=1$, and all its
derivatives at the origin vanish: however it goes to zero sufficiently
fast at infinity. Since $k/n$ is conjugate to the angle $\theta$, we
should think of this cut-off as being equivalent to a discretisation
$\delta\theta$. Thus we should choose $\Lambda_n=\Lambda\cdot n$, where
$\Lambda\sim(\delta\theta)^{-1}$.

To perform the sum over $k$, we use
the Euler-MacLaurin (EML) sum formula \cite{as}
\be
\frac{1}{2}\sum_{k=0}^\infty d_k f(k)= \int_0^\infty f(k) dk
- \frac{1}{12}f'(0)-
\sum_{j=2}^\infty \frac{B_{2j}}{(2j)!} f^{(2j-1)}(0)\,,
\label{eml}
\ee
where $B_{2n}$ are the Bernoulli numbers \cite{as}.
Using standard identities of the Bessel function (namely
$\partial_k K_k(x)|_{k=0}=0$ and
$\partial_k I_k(x)|_{k=0}=-K_0(k)$ \cite{as}), we obtain
\be
G_n(r)=
2 \int_0^\infty dk I_{k/n} (m r) K_{k/n} (mr)F(k/n\Lambda)+\frac{1}{6n} K_0^2 (mr)\,
+\int_0^\infty r dr
\sum_{j\geq 1} \frac{B_{2j}}{(2j)!n^{2j+1}} D_{2j +1}(r)\,,
\label{Greml}
\ee
where we define
$D_i(x)=\partial^i(I_k(x) K_k(x))/\partial k^i|_{k=0}$.
In the last term in Eq. (\ref{Greml})
the order of the integral, derivative, and sum can be
exchanged and each term in the sum is
\be
\frac{\partial^i}{\partial k^i} \int_0^\infty x dx I_k(x) K_k(x)=
-\frac{\partial^i}{\partial k^i} \frac{k}{2}=0 \quad {\rm for }\, i=2j+1\geq2\,,
\ee
i.e. the sum in Eq. (\ref{Greml}) is vanishing.

Thus, still with the regulator in place,
\be
\frac{\partial}{\partial m^2} \log Z_n=
-\int_0^\infty r dr \int_0^\infty dk I_{k/n} (m r) K_{k/n} (mr)F(k/n\Lambda)
-\frac1{24nm^2}
\ee
where we have used $\int_0^\infty rK_0^2(mr)dr=1/(2m^2)$.

Now the point is that in the integral over $k$ the factor of $n$ can be
scaled out by letting $k\to nk$. Thus this potentially divergent term
cancels in the required combination $G_n-nG_1$. Having taken this
combination, we may now remove the regulator to find the main result
\be
\frac{\partial}{\partial m^2} \frac{\log Z_n}{Z^n}=
\frac{1}{24 m^2}\left(n-\frac{1}{n}\right)\,.
\ee
The integration of the last expression wrt $m^2$ (made following the recipe
for the integration limits given in the previous section) gives
\be
\log {\rm Tr} \rho^n= \log\frac{Z_n}{Z^n}=
\frac{\log a^2 m^2}{24}\left(n-\frac{1}{n}\right)\,,
\label{trrhon}
\ee
and finally the entanglement entropy
\be
S=-{\rm Tr} \rho\log\rho=
-\left. \frac{\partial}{\partial n} {\rm Tr} \rho^n\right|_{n=1}=
-\left. \frac{\partial}{\partial n}
(m^2 a^2)^{\frac{1}{24}\left(n-\frac{1}{n}\right)}
\right|_{n=1}
=-\frac{1}{12} \log m^2 a^2\,,
\ee
that agrees with the general formula we derive (\ref{mass}), with $c=1$ and
$m=\xi^{-1}$.

\subsection{Integrable models and the corner transfer matrix}

In this subsection we verify (\ref{mass})
for the transverse Ising chain and the uniaxial
XXZ Heisenberg model.
These results are also an
independent check of the uniform convergence of the derivative wrt $n$
of $\rho^n$ when $n\rightarrow1$, since for these models we can compute
the eigenvalues of $\rho_A$ exactly.

Although these systems are integrable and their ground state is known, a
direct calculation of $\rho$ is difficult. The difficulties arising in a
direct calculation can be avoided
mapping the quantum chains onto two-dimensional classical spin systems.
As firstly pointed out by Nishino et al. \cite{nishino}
the density matrix of the
quantum chain is the partition function of a two-dimensional strip with a cut
perpendicular to it. In fact the ground state of a quantum chain described
by an Hamiltonian $H$ is also eigenstate of the transfer matrix $T$ of a
classical systems satisfying $[H,T]=0$.
Therefore the reduced density matrix of a subsystem $A$ of the chain (defined
in the Introduction as
$\rho_A={\rm Tr}_B |\Psi\rangle\langle \Psi|$, with
$B$ the complement of $A$) is the
partition function of two half-infinite strips, one extending from $-\infty$
to $0$ and the other from $+\infty$ to 0, with the spins in $B$
identified.

This partition
function is the product of four Baxter corner transfer matrices
(CTMs)\cite{Baxter}
$\hat A$.
If the lattice is choose in a clever way (i.e. rotated by $\pi/4$ with
respect to the cut)
one ends in the infinite length limit with \cite{pkl-99}
\be
\hat{\rho}_A=\hat{A}^4=e^{-\hat{H}_{\rm CTM}}\,,
\ee
where $\hat{H}_{\rm CTM}$ is an effective Hamiltonian, which, for the
models under consideration, may be diagonalised by means
of fermionisation (see for more details about this equivalence
\cite{pkl-99} and references therein.)
Note that ${\rm Tr \hat{\rho}_A \neq 1}$, thus the usual density matrix is
$\rho_A=\hat{\rho}_A/{\rm Tr}\hat{\rho}_A$.

The method just outlined is very general. However, for the
integrable chains under consideration (and indeed for any model
satisfying suitable Yang-Baxter equations\cite{Baxter}) it is
possible to  write $H_{\rm CTM}=\e \hat{O}$, with $\e$ the scale giving the
distance between the energy levels, and $\hat{O}$ is an
operator with integer eigenvalues (for the Ising and XXZ models it is
expressed in terms of free fermions.)
Using this property the entropy is given by\footnote{In the rest of
this section, all logarithms are assumed taken to base $e$.}
\be
S=-{\rm Tr} \rho_A\log\rho_A=
-{\rm Tr} \frac{\hat{\rho}_A \log\hat{\rho}_A}{{\rm Tr}\hat{\rho}_A}+
\log {\rm Tr}\hat{\rho}_A=
-\e\frac{\partial\log Z}{\partial \e}+ \log Z\,,
\label{entrint}
\ee
where we defined $Z={\rm Tr} \hat{\rho}_A={\rm Tr} e^{-\hat{H}_{\rm CTM}}$.

The Ising model in a transverse field can be described by the one-dimensional
Hamiltonian
\be
H_I=-\sum_{n=1}^{L-1}\sigma^x_n-
\lambda  \sum_{n=1}^{L-1}\sigma^z_n\sigma^z_{n+1}\,,
\label{HamI}
\ee
where $\sigma_n$ are the Pauli matrices at the site $n$, and
we normalise the Hamiltonian (following \cite{pkl-99}) by imposing the
transverse field in the $x$ direction to be one, thus the transition is
driven by the parameter $\lambda$.
The classical equivalent of (\ref{HamI}) is two dimensional Ising model.
For $\lambda=0$ the ground state of the Hamiltonian (\ref{HamI}) is a quantum
``paramagnet'' with all the spins aligned with the magnetic field in the
$x$ direction, and $\langle\sigma^z_n\rangle=0$. In the opposite limit
$\lambda=\infty$ the magnetic field is negligible and the ground state is
ferromagnetic with $\langle\sigma^z_n\rangle=\pm 1$.
The (second-order) transition between this two regimes happens at $\lambda=1$.
The exponent characterizing the divergence of the correlation length
is $\nu=1$, i.e. $\xi\simeq |\lambda-1|^{-1}$.

The CTM of the Ising model may be diagonalised in terms of fermionic
operators. The CTM Hamiltonian, written in terms of the fermion occupation
number $\hat{n}_i$ (with eigenvalues $0$ and $1$) is \cite{pkl-99}
\be
\hat{H}_{\rm CTM}=\sum_{j=0}^\infty \e_j \hat{n}_j\,.
\ee
The energy levels are
\be
\e_j=
\left\{\matrix{
(2j+1)\e &\quad {\rm for} \,\lambda<1\,,\cr
2j\e     &\quad {\rm for} \,\lambda>1\,,
}\right.
\quad {\rm with}\;
\e=\pi \frac{K(\sqrt{1-k^2})}{K(k)}\,
\ee
where $K(k)$ is the complete elliptic integral of the first kind \cite{as},
and $k=\min [\lambda,\lambda^{-1}]$.

\begin{figure}[t]
\centerline{\epsfig{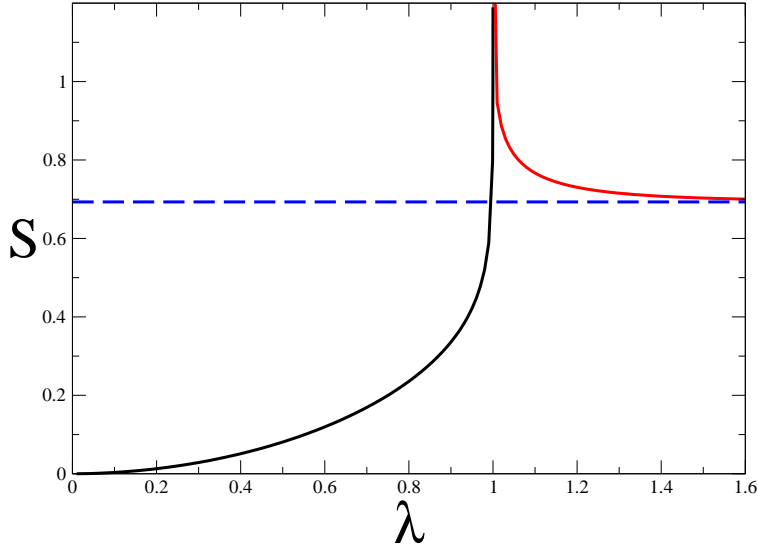}}
\caption{Entanglement entropy for the 1D Ising chain as function of $\lambda$.
The dashed line is the limit for $\lambda\rightarrow\infty$, i.e. $\log 2$.}
\label{Fig}
\end{figure}

For $\lambda<1$
\be
Z={\rm Tr} e^{-\hat{H}_{\rm CTM}}=\prod_{j=0}^\infty\left[1+e^{-\e(2j+1)}\right]\,,
\ee
and the entropy from Eq. (\ref{entrint}) is
\be
S=
\e \sum_{j=0}^\infty \frac{2j+1}{1+e^{(2j+1)\e}}+
\sum_{j=0}^\infty\log (1+e^{-(2j+1)\e})\,.
\label{Spara}
\ee
Analogously in the quantum ferromagnetic phase ($\lambda>1$)
\be
S=\e \sum_{j=0}^\infty \frac{2j}{1+e^{2j\e}}+
\sum_{j=0}^\infty\log (1+e^{-2j\e})\,.
\label{Sferro}
\ee
Fig. \ref{Fig} shows a plot of the entropy as function of $\lambda$,
characterised by a divergence at the quantum critical point $\lambda=1$.
Note that $S(0)=0$ and $S(\infty)=\log2$, in agreement with the expectation
that the pure ferromagnetic ground state ($\lambda=\infty$) has two possible
accessible configurations with opposite sign of magnetisation (i.e.
$S(\infty)=\log2$) and the pure quantum paramagnetic ground
state ($\lambda=0$)
has only one configuration available with all the spins aligned in the
direction of the magnetic field $x$ and the resulting entropy is zero.

Let us analyze in details the behaviour at the critical point.
For $\lambda\rightarrow1$, $\e\rightarrow0$ in both the phases, thus
the sums Eqs. (\ref{Spara}) and (\ref{Sferro}) can be approximated by the
integral
\be
S \simeq  \int_0^\infty dx \left(\frac{2x\e}{1+e^{2x\e}}+\log (1+e^{-2x\e})
\right)
=\frac{\pi^2}{12} \frac{1}{\e}\,,
\label{Sint}
\ee
with $\simeq$ we mean in the critical region.
The $\lambda$ dependence (we use $K(0)=\pi/2$,
$K(x)=-1/2 \log(1-x) +O((1-x)^0)$ \cite{as})
\be
S \simeq  \frac{\pi^2}{12} \frac{1}{\e}
\simeq-\frac{1}{12}\log (1-k)=\frac{1}{12}\log \xi+C_1\,,
\ee
where in the last equality $\xi\propto |1-k|^{-1}$ has been used.
This agrees with \ref{mass}, with $c=1/2$ for the Ising model.
The constant $C_1$ is not universal.


Another model whose density matrix has been derived by using the CTM is the
XXZ model
\be
H_{XXZ}=\sum_n (\sigma^x_n\sigma^x_{n+1}+\sigma^y_n\sigma^y_{n+1}+
\Delta \sigma^z_n\sigma^z_{n+1})\,,
\ee
for $\Delta>1$ (whose classical equivalent is the Baxter six-vertex
model \cite{Baxter}.)
This model has an Ising-like ferromagnetic state for $\Delta>1$ and a planar
XX ferromagnetic ground state at $0<\Delta<1$ (the case $\Delta<0$ describes
the antiferromagnetic regime, in which we are not interested.)
At $\Delta=1$ the Hamiltonian is isotropic (XXX) and approaching such a point
from large $\Delta$ values the correlation length diverges as
(see e.g. \cite{Baxter})
\be
\log \frac{\xi}{a}\simeq\frac{\pi^2}{2\sqrt2\sqrt{\Delta-1}}\,,
\label{xiXXZ}
\ee
because at $\Delta=1$ a massless excitation (Goldstone mode)
is present in the spectrum.

The CTM Hamiltonian $\hat{H}_{\rm CTM}$ for this model has been obtained
in Ref. \cite{pkl-99}
\be
\hat{H}_{\rm CTM}=\sum_{j=0}^\infty 2j \e n_j\,,
\ee
with $\e={\rm arccosh} \Delta$, for $\Delta>1$.
The entropy is given by Eq. (\ref{Sferro}).
Close to the isotropic point ($\Delta=1$) it holds
$\e\simeq\sqrt2\sqrt{\Delta-1}$ and (see Eq. (\ref{Sint}))
\be
S\simeq\frac{\pi^2}{12}\frac{1}{\e}\simeq\frac{\pi^2}{12\sqrt2}
\frac{1}{\sqrt{\Delta-1}}\,.
\ee
Using Eq. (\ref{xiXXZ}) to write $\Delta$ in term of the correlation length,
we have
\be
S\simeq \frac{1}{6} \log \frac{\xi}{a}\,,
\ee
in agreement with \ref{mass}, with $c=1$.
Again, in the limit $\Delta\rightarrow\infty$, $S=\log2$, since the
ferromagnetic ground state is Ising-like.

The method used here for the transverse Ising model and the XXZ
model can in principle be applied  to all those integrable model
whose weights satisfy Yang-Baxter equations.

\section{Finite-size effects}
\label{sectfss}

So far, we have studied the entropy either at the critical point, for a
finite subsystem, or away from criticality in an infinite system. These
two regimes may be linked by a generalisation of finite-size scaling
theory.
This would assert, for example,
that the entropy of a subsystem $A$ which forms e.g. the left half of a
finite non-critical system of length $2L$ should have the form
\be
S_A(L,\xi)= \frac{c}{6}\left(\log L +
s_{\rm FSS}(L/\xi)\right)\,,
\label{FSS1d}
\ee
with $s_{\rm FSS}(0)=0$ (this because we are referring to the scaling part,
the constant term found before is not universal and it defines an overall
additive normalisation of $S$) and
$s_{\rm FSS}(x)\sim-\log x$, for large $x$, so as to recover $S=(c/6)\log\xi$
in the infinite length limit.
$s_{\rm FSS}(x)$ should admit a small $x$ expansion
\be
s_{\rm FSS}(x)=\sum_{j\geq1} s_j x^{2j}\,,
\ee
with $s_j$ {\it universal} coefficients, and a large $x$ expansion
\be
s_{\rm FSS}(x)=-\log x+\sum_{j\geq0} \frac{s^\infty_j}{x^{j}}\,,
\ee
with $s^\infty_j$ also universal (in special situations
logarithmic corrections could also be generated.)
In the next section we test this hypothesis for the free massive field
theory.

\subsection{FSS in the Gaussian model}

Once again we consider the $n$-sheeted surface, but it now consists
of $n$ discs of finite radius $L$, sewn together along the negative real
axis from $-L$ to $0$.
The Gaussian two-point function in a finite geometry
has been calculated in the Appendix Eq. (\ref{2ptL}).
Setting $\theta=\theta'$, $r=r'$ and integrating over $\theta'$
the propagator at coincident point is
\be
G_n(r)= \sum_{k=0}^\infty d_k \sum_{i=1}^\infty
\frac{2/L^2}{J^2_{k/n+1}(\alpha_{k/n,i})}
\frac{J_{k/n}^2(\alpha_{k/n,i} r/L)}{\alpha_{k/n,i}^2/L^2+m^2}\,,
\ee
where $\alpha_{\nu,i}$ denotes the $i-$th zero of $J_\nu(x)$.
$G(r)$ is the analogous of (\ref{Gr}) in a finite geometry.

As we showed in the previous section, the right order to proceed to get a
sensitive $Z_n$ is first to perform the sum over $i$,
then sum over $k$ and finally integrate over $r$.
This is really hard, requiring a sum of Bessel functions
for generic argument, over the zeroes of different Bessel functions.
What one can do is inverting the order of the sums and integrations and try
to understand what happens. From the previous exercise
we know that this operation has to be done with care.

However, the formal result is
\be
\log Z_{n}= \frac{1}{2}
\sum_k d_k\sum_i\log\frac{\alpha^2_{k/n,i}/L^2+m^2}{\alpha^2_{k/n,i}/L^2+a^{-2}}.
\label{formal}
\ee
Since the large $i$ behaviour of the zeroes of the
Bessel functions is $\alpha_{\nu,i}\sim\pi(i+\nu/2-1/4)$ (see e.g. \cite{as}),
this sum diverges.

A first, qualitatively correct expression for the
universal function $s_{\rm FSS}(x)$ can be obtained from Eq. (\ref{formal}),
truncating the EML formula Eq. (\ref{eml}) in the variable $k$ at
the first order in the derivative.
This approximation gives a correct form of the result because it takes into
account the complete infinite volume result (whose truncated expression
is exact.) The integral term in the EML is divergent, but it cancels as
before in the
ratio $Z_n/Z^n$ if the cut-off in the angular modes is properly chosen as in
the infinite volume case.
The EML approximation at the first order is (we do not write the integral)
\be
\log Z_{n}=
\frac{\pi L^2 (a^{-2}-m^2)}{12n} \sum_i
\frac{\alpha_{0,i}}{(\alpha_{0,i}^2+m^2L^2)(\alpha_{0,i}^2+L^2/a^2)}
\frac{Y_0(\alpha_{0,i})}{J_1(\alpha_{0,i})}\equiv\frac{F_1(mL,L/a)}{n}\,,
\label{Zeml1}
\ee
where we used
\be
\left.\frac{\partial \alpha_{k,i}}{\partial k}\right|_{k=0}=
\frac{\pi}{2}\frac{Y_0(\alpha_{0,i})}{J_1(\alpha_{0,i})}\,,
\ee
which can be derived using the formulas \cite{as} of the derivative of the
Bessel functions with respect to the order.
The remaining sum over $i$ (the zeroes of the Bessel function) is now finite.

This gives the entropy as
\be
S=-\left. \frac{\partial}{\partial n}{\rm Tr} \rho^n\right|_{n=1}=
2 F_1(m L,L/a)\,,
\ee
which reproduces the correct limits for $m=0$ and $L=\infty$.

From this formula we may compare with the FSS ansatz (\ref{FSS1d}),
with $c=1$:
\be
s_{\rm FSS}^{(1)}(m L)=12 (F_1(m L,L/a)- F_1(0,L/a))=
-{\pi} \sum_i
\frac{Y_0(\alpha_{0,i})}{J_1(\alpha_{0,i})}
\frac{m^2 L^2}{\alpha_{0,i}(\alpha_{0,i}^2+m^2L^2)}\,,
\label{sfs}
\ee
that, as expected, is a function only of the product $mL$ and all the
dependence upon $a$ disappears after the subtraction.
This agrees with $s_{\rm FSS}(0)=0$ and
$s_{\rm FSS}(0)=-\log x$, for large $x$ (it can be easily shown, since in
this limit the sum can be replaced by an integral.)
The over-script $(1)$ is there to remind that we made a first-order
approximation in EML.

Eq. (\ref{sfs}) characterises completely the crossover between the
mass dominated (non-critical) and the geometry dominated (critical) regime.

\begin{figure}[t]
\centerline{\epsfig{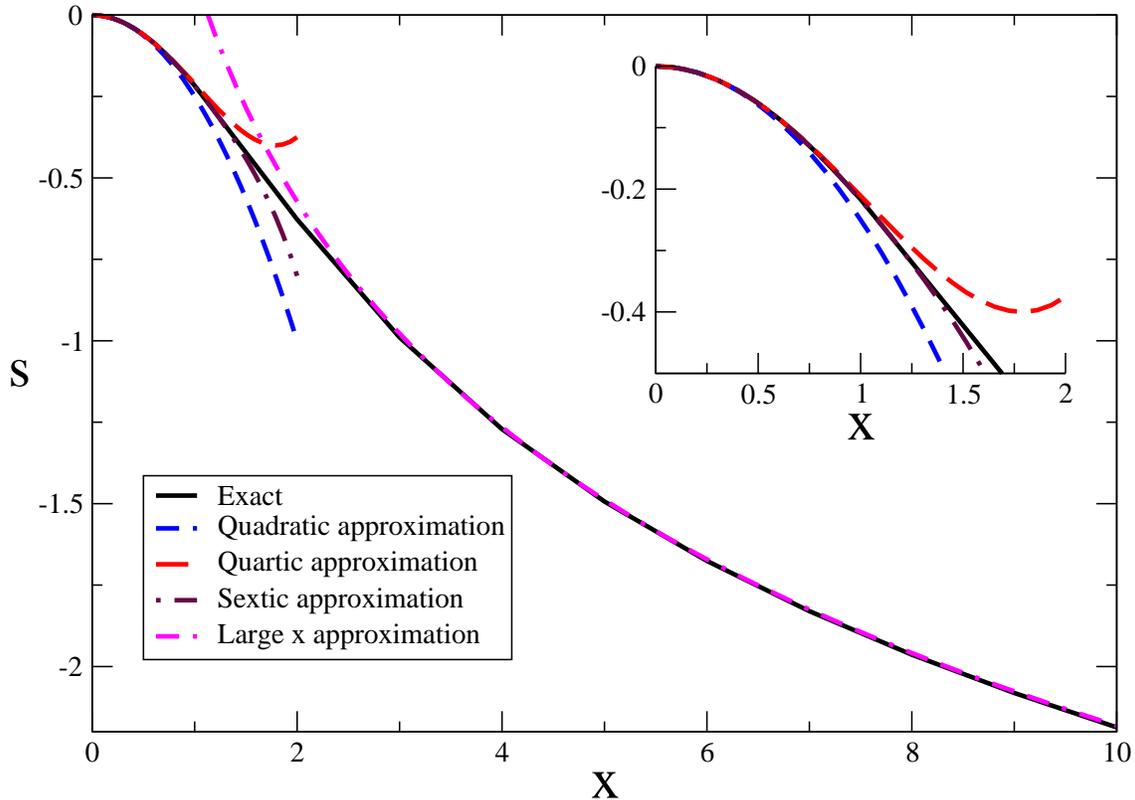}}
\caption{
$s_{\rm FSS}^{(1)}(x)$: Exact expression obtained as numerical sums
over the first 1000 zeroes of the Bessel functions compared with
small (quadratic, quartic and sextic) and large $x$ ($\log x+s_0^\infty$)
approximations.
Even the use of quadratic and large $x$ approximation may reproduce the
right formula over all the range.
Inset: Comparison of several small $x$ approximants for $x<2$.
}
\label{figs}
\end{figure}

A plot of $s_{\rm FSS}^{(1)}$ is shown in Fig. \ref{figs} (calculated
as the sum of the first 1000 zeroes of  $J_0$.)
From this figure we see that an optimum approximation for $x\geq2$ of
$s_{\rm FSS}^{(1)}(x)$ is $s_{\rm FSS}^{\rm asy}(x)= -\log x+s_0^\infty$,
with $s_0^\infty=0.120912$ (this value of $s_0^\infty$ is a fit.)
We can also calculate analytically all the universal
coefficients of the small $mL$ expansion \cite{bessels}:
\bea
s_1^{(1)}&=-&
{\pi}\sum_{i=1}^\infty \frac{Y_0(\alpha_{0,i})}{J_1(\alpha_{0,i})}
\frac{1}{\alpha_{0,i}^3}=-\frac{1}{4}\,,\\
s_2^{(1)}&=&
{\pi}\sum_{i=1}^\infty \frac{Y_0(\alpha_{0,i})}{J_1(\alpha_{0,i})}
\frac{1}{\alpha_{0,i}^5}=\frac{5}{128}\,,\\
s_3^{(1)}&=-&
{\pi}\sum_{i=1}^\infty \frac{Y_0(\alpha_{0,i})}{J_1(\alpha_{0,i})}
\frac{1}{\alpha_{0,i}^7}=-\frac{23}{3456}\,,\\
s_4^{(1)}&\simeq& 1.15 \times 10^{-3},\quad
s_5^{(1)}\simeq -1.98 \times 10^{-4},\quad
s_6^{(1)}\simeq3.42 \times 10^{-5}.
\eea
Fig. \ref{figs} (see also the inset) shows a plot of $s_{\rm FSS}^{(1)}(x)$,
compared with quadratic, quartic, and sextic approximation.
The agreement is excellent in the region $x\leq2$.

Before starting the full calculation of the function $s_{\rm FSS}(x)$
let us summarise what we can learn from the first order approximation in
the EML expansion: A rather good approximation of the full function
may be obtained by matching only the first-order small argument expansion
with  the large $x$ behaviour $\sim\log x$
(which is exact and does not depend upon
the approximation.) Thus in the following we show how to calculate the
coefficients $s_j$ without any approximation.

Our starting point is again Eq. (\ref{formal}), but this time we will not
use EML sum formula. For this reason we have to be care since the sum is
divergent.

The interesting object is $\log Z_{n}-n\log Z$.
In particular we can write an FSS ansatz also for this ``free-energy''
and so subtract the $m=0$ part, obtaining the universal function $f(x=mL)$
\be
f(x)=3\sum_k d_k\sum_i\left[ n\log\left(1+\frac{x^2}{\alpha_{k,i}^2}\right)
-\log\left(1+\frac{x^2}{\alpha_{k/n,i}^2}\right)\right]\,,
\ee
form which
$s_{\rm FSS}(x)=-\left.\frac{\partial f(x)}{\partial n}\right|_{n=1}$.
This formula has to be intended in a formal way: in fact it is the difference
of two diverging sums. To make this difference sensible the recipe for
the cut-off  explained in the previous section has to be used.
Anyway from the computational point of view it is simpler to make the
calculation without care of the cut-off and to adjust only at the end the
result.
The numerical sum over the zeroes of the Bessel functions of generic order
cannot be done as before, but the small $x$ expansion of
$f(x)=\sum f_jx^{2j}$ with coefficients
\be
f_j=\frac{(-1)^{j+1} 3}{j} \sum_k d_k\sum_i\left[ \frac{n}{\alpha_{k,i}^{2j}}-\frac{1}{\alpha_{k/n,i}^{2j}}\right]\,,
\ee
can be calculated analytically. For example
\be
s_1^{\rm wrong}=-\left.\frac{\partial}{\partial n} f_1 \right|_{n=1}=
3\sum_k d_k\sum_i \frac{1}{\alpha_{k,i}^2}\left[-1+\frac{2k}{\alpha_{k,i}}
\frac{\partial\alpha_{k,i}}{\partial k}\right]=
-3\sum_k d_k  \frac{\partial}{\partial k}k\sum_i\frac{1}{\alpha_{k,i}^2}\,,
\ee
that using \cite{bessels}
\be
\sum_{i=1}^\infty\frac{1}{\alpha_{k,i}^2}=\frac{1}{4(k+1)}\,,
\ee
leads to
\be
s_1^{\rm wrong}=-3\sum_k d_k  \frac{\partial}{\partial k} \frac{k}{4(k+1)}=
-\frac{3}{4}\sum_k d_k \frac{1}{(k+1)^2}=
-\frac{3}{4}(2\frac{\pi^2}{6}-1)\,.
\label{app}
\ee
We use the over-script ``wrong'' because it is not strictly correct to make the
calculation this way, since we are implicitly using the same cut-off for $Z$
and $Z_n$.
In fact, applying the EML to the sum in (\ref{app}), one has
\be
s_1^{\rm wrong}=-\frac{3}{4}\sum_k d_k \frac{1}{(k+1)^2}=
-\frac{3}{4}\left(2\int_0^\infty \frac{dk}{(k+1)^2}+\frac{1}{6} 2+\dots\right)\,,
\ee
that has a first order term $s^{(1)}_1=-1/4$ (in agreement with what
previously found) but the integral is not vanishing.
This is a finite difference of the two divergent expressions with the
wrong cut-off.
The value of the integral ($2\int dk (k+1)^{-2}=2$)
must be properly subtracted to have the
right result
\be
s_1=-\frac{3}{4}(2\frac{\pi^2}{6}-3)=-0.217401\dots\,.
\ee
This is close to the first order EML result $-1/4$, signaling that
such approximation, not only reproduces the qualitative physics but is
also quantitatively reliable (at the level of 10\%.)
In the same manner one can calculate all the
coefficients $s_j$, using the more complicated expressions for the sum of
higher negative powers of zeroes of Bessel functions reported in the
literature \cite{bessels}.

\section{Higher dimensions: Scaling of entropy and area law}
\label{secthd}

The scaling hypothesis plays a fundamental rule in understanding
classical phase transitions. Crudely speaking, it asserts that the
microscopic length scale $a$ does not enter explicitly into thermodynamic
relations near the critical point, as long as the various thermodynamic
variables are suitable normalised and
allowed to scale with their non-trivial scaling dimensions, related to
the various universal critical exponents.

In fact, from the scaling of the singular part of the free energy density
(here $t=|T-T_c|/T_c$, $\xi\propto t^{-\nu}$ is the correlation length,
$h$ the external magnetic field, $f_{\pm}(x)$ a universal function,
the subscript $\pm$ refers to the two phases, and $y_h$ is the scaling
dimension of $h$)
\be
f_{\rm sing}(t,h)=\xi^{-d} f_{\pm}(h\xi^{-y_h})\,,
\ee
the critical behaviour of all the thermodynamic observables and in particular
the scaling laws may be derived. Note that $f_{\rm sing}$ is not the
total free energy density: there is another non-universal piece which
has an explicit $a^{-d}$ dependence on the microscopic cut-off. However,
this term is analytic in the thermodynamic variables.

As argued by Srednicki\cite{s-93}, for $d>2$ the entropy $S_A$ is
proportional to the surface area $\cal A$ of the subsystem $A$. Thus we
should discuss the entropy per unit area $s=S_A/{\cal A}$.
In analogy with the classical case, we may conjecture a scaling
form for the singular part of the
entropy per unit area near a quantum phase transition
\be
s_{\rm sing}(g,h,T)=\xi^{-(d-1)}s_\pm(h\xi^{-y_h},T \xi^{-z})\,,
\label{Sscal}
\ee
where $y_h>0$ describes the relevant effect of the field conjugated to the
order parameter, $z>0$ the
relevance of the temperature close the the quantum phase transition (in all
the examples considered up to now $z=1$), and all
other neglected operators are supposed to be irrelevant.
$s_\pm$ should be a universal function apart from
the normalisations of its arguments.
The relation between $\xi$ and $g$ (the parameter driving the transition)
is $\xi=|g-g_c|^{-\nu}$.
As for the free energy, $s_{\rm sing}$ is not the total entropy: we also
expect to find explicitly $a$-dependent pieces which are, however,
analytic in $g$.

From the scaling of the entropy all the scaling laws
can be obtained, e.g. the specific heat goes like
\be
C=T\frac{\partial S}{\partial T}\sim T \xi^{-(d-1+z)}\,.
\ee
Similar identities can be derived for other observables.

This scaling can be explicitly checked in the case of the Gaussian
model in all dimensions for $h=T=0$ and $z=1$.
To calculate the entropy of a $d$-dimensional system
close to a quantum critical point, one has to consider a $d+1$ field theory.
We consider the geometry where $d-1$ dimensions are translational invariant
in a domain delimited by an hypersurface of area $\cal A$,
and on the remaining two dimensional plane there is a branch cut going
from 0 to $\infty$ in an arbitrary direction.
The choice of a different geometry is not expected to change the main results.

The two-point function is translational invariant in $d-2$ directions and
coincides with Eq. (\ref{2pt}) on the cut plane, i.e.
\be
G(r_\perp;r,\theta,r',\theta')=
\int \frac{d^{d-1} k_\perp}{(2\pi)^{d-1}} e^{i k_\perp r_\perp}
\frac{1}{2\pi n} \sum_{k=0}^\infty d_k \int_0^\infty \lambda d\lambda
\frac{J_{k/n}(\lambda r)J_{k/n}(\lambda r')}{\lambda^2+m^2+k_\perp^2}
{\cal C}_k (\theta,\theta')\,,
\label{2pthd}
\ee
where $r_\perp$ is the vector between the two considered points in the
$d-1$ dimensional subspace.

The calculation proceeds as in the one-dimensional case with the substitution
$m^2\rightarrow m^2+k_\perp^2$, and the integration over $k_\perp$ in front
of all the equations. The space integration is obviously over
$d^{d-1}r'_\perp r dr d\theta$ (we use $r'_\perp$ to stress that it is
different from that appearing in Eq. (\ref{2pthd}) that instead is
$r_\perp=0$, since the two-point function at coincident point has to be
considered.)
Thus the analogue of Eq. (\ref{trrhon}) is
\be
\log {\rm Tr} \rho^n=
\int d^{d-1} r'_\perp \int \frac{d^{d-1}k_\perp}{(2\pi)^{d-1}}
\frac{\log (k_\perp^2+m^2)}{24}\left(n-\frac{1}{n}\right)\,.
\ee
To have a finite result both the integrations are in a finite region.
This means that the system must live in a finite box with
$\int d^{d-1} r'_\perp={\cal A}$ and a cut-off proportional to $a^{-1}$
must be understood for the integration over $k_\perp$.

In this way the entropy is
\be
S
=-\frac{\cal A}{12} \int \frac{d^{d-1}k_\perp}{(2\pi)^{d-1}}
\log\frac{k_\perp^2+ m^2}{k_\perp^2+ a^{-2}}\,.
\label{Sd}
\ee
In general, the integral diverges like $a^{-(d-1)}$. However the
coefficient of this divergence (and other sub-leading divergences which
can occur for sufficiently large $d$) are in general analytic in $m$.
There is, however, a finite piece which behaves as $m^{d-1}$ in
agreement with (\ref{Sscal}). Note however that as $d\to1$ this singular
term combines with the non-singular piece $\propto a^{-(d-1)}$ to give
the previously found result $\propto\log(ma)$.

Just as for the classical free energy, one may also conjecture a finite-size
scaling form for the entropy of the form
\be
s_{\rm sing}(L,g,h,T)=L^{-(d-1)}
s_{\rm FSS}(L^{1/\nu} |g-g_c|, L^{y_h} h,L^{z} T)\,.
\ee
For $h=T=0$ it reads
\be
s_{\rm sing}(L,g,h=T=0)= A_s L^{-(d-1)}  s_{\rm FSS}(L/\xi)\,,
\ee
with $s_{\rm FSS}(L/\xi)$ a function which satisfies $s_{\rm FSS}(0)=1$ and
 $s_{\rm FSS}(x)\propto 1/x$ for large $x$, in order to reproduce the
infinite volume limit, and $A_s$ is a non universal constant fixing the
normalisation of the entropy.
In one dimension, this relation again does not make sense. The $\log L$
behaviour may be seen as resulting from a cancellation between the
$L^{-(d-1)}$ behaviour of $s_{\rm sing}$ against a constant term coming
from the regular part, with $A_s=O((d-1)^{-1})$ as $d\to1$.
This gives the anticipated form
\be
S_{\rm sing}(L,\xi)= \frac{c}{6}\left(\log L +
s_{\rm FSS}(L/\xi)\right)\,,
\ee
that has been explicitly checked for the Gaussian model in Sec. \ref{sectfss}.

\vspace{1cm}

\noindent\em Acknowledgments\em. JC thanks D.~Huse for first bringing
this problem to his attention.
This work was supported in part by the EPSRC under Grant GR/R83712/01
The initial phase was carried out while JC was a member of the Institute
for Advanced Study.
He thanks the School of Mathematics and the School of Natural Sciences
for their hospitality.
This stay was supported by the Bell Fund, the James D.~Wolfensohn Fund,
and a grant in aid from the Funds for Natural Sciences.

\section{Note added.}

We recently were made aware, some time after this paper appeared
in print, that the result (3.32) for more than one interval is
incorrect in general. This error can be traced to the assumption
at the beginning of Sec.~3C that there exists a uniformizing
transformation for the $n$-sheeted Riemann surface into the
sphere. In fact the surface has in general a non-trivial genus and
therefore such a transformation does not exist. For the case of
two intervals and $n=2$ it can readily be seen that the surface is
topologically a torus, and therefore the partition function $Z_2$
depends on the whole operator content of the theory, not only the
central charge as claimed. For other values of $n$ the partition
function should also depend on the structure constants (OPE
coefficients). It is therefore highly unlikely that this would not
remain true for the derivative at $n=1$. Indeed, a study of the
computation for the 2-interval case for a compactified boson,
first carried out in 1987 by Dixon et al.\cite{Dixon}, shows that
this is the case, and that our (3.32) is incorrect. This is also
the result of recent numerical investigations
\cite{Gliozzi,Furokawa}. We are grateful to V.~Pasquier and
F.~Gliozzi for pointing out this work.

Although this error also carries over to the case of a
semi-infinite system, mentioned just below (3.32) and also to
Eq.~(3.33), it does not impact on our later application of these
methods to studying the time-dependence of the entanglement
entropy following a quantum quench\cite{cc3}. This is because in
this case only the limiting behavior, when the various
cross-ratios are either small or large, is needed, and this is
insensitive to the precise form of (3.32). Likewise, as far as we
are aware, all the other conclusions of the present paper remain
valid.

\appendix

\section{Two-point function in a $n$-sheeted Riemann geometry:
Infinite and finite volume results}

The Green function $G({\bf r,r'})$ of the Helmholtz differential equation
$(-\nabla^2_{\bf r}+m^2)f({\bf r})=0$, with specific boundary condition,
is the solution of
$$
(-\nabla^2_{\bf r}+m^2)G({\bf r,r'})=\delta^d({\bf r-r'})\,.
$$
$G({\bf r,r'})$ admits an eigenfunction expansion
$$
G({\bf r,r'})=\sum_k N_k \phi_k({\bf r}) \phi_k({\bf r'})\,,
$$
in terms of $\phi_k({\bf r})$, eigenfunctions of the Helmholtz differential
operator $(-\nabla^2_{\bf r}+m^2)$.
$N_k$ is a  normalisation constant that should be derived from the
orthonormality requirement of the eigenfunctions
$$
N_m\int \phi_m({\bf r})\phi_n({\bf r'})d^d\!r=\delta_{nm}\,.
$$

In the case of a $n$-sheeted Riemann surface we are interested in,
the eigenvalue problem is solved in polar coordinates
${\bf r}=(x,y)=(r\cos\theta,r\sin\theta)$.
A complete set of eigenfunctions is
\be
\phi_{\nu,i}^a=\cos({i\nu \theta}) J_\nu(\lambda_i r)\,,\qquad
\phi_{\nu,i}^b=\sin({i\nu \theta}) J_\nu(\lambda_i r)\,,
\ee
with $J_\nu(x)$ Bessel functions of the first kind.
Note that the Bessel functions of the second kind $Y_\nu(x)$, that are also
eigenfunctions of the same differential operator, do not enter in
the expansion since we require regularity at $r=0$.

Let us consider first the solution in a finite geometry.
The infinite volume limit is recovered by taking the limit
$L\rightarrow \infty$ in the sense of distributions.
Imposing the $2\pi n$ periodicity boundary condition we have that $\nu$ is
an integer multiple of $1/n$, i.e. $\nu=k/n$.
Constraining the eigenfunctions to vanish at $r=L$, the eigenvalues
are $\lambda_i L=\alpha_{\nu,i}$, with $\alpha_{\nu,i}$ the $i$-th zero of
the Bessel function.

Using the orthogonality relation of the Bessel functions (see e.g. \cite{as})
$$
\int_0^L r dr J_\nu(\alpha_{\nu,i} r/L)J_{\nu'}(\alpha_{\nu',i} r/L)=
\frac{L^2}{2}J^2_{\nu+1}(\alpha_{\nu,i})\delta_{\nu\nu'}\,,
$$
we get from the orthonormality requirement
$$
N_{i,k}=\frac{d_k}{2\pi n}\frac{2/L^2}{J^2_{k/n+1}(\alpha_{k/n,i})}\,.
$$
that gives the Green function
\be
G(r,\theta,r',\theta')=
\frac{1}{2\pi n} \sum_{k=0}^\infty d_k \sum_{i=1}^\infty
\frac{2/L^2}{J_{k/n+1}^2(\alpha_{k/n,i})}
\frac{J_{k/n}(\alpha_{k/n,i}r/L)J_{k/n}(\alpha_{k/n,i}r'/L)}{\alpha_{k/n,i}^2/L^2+m^2}
{\cal C}_k(\theta,\theta')\,.
\label{2ptL}
\ee
The factor $d_k$ comes from $\int_0^{2\pi} 1dx =2\pi$ and
$\int_0^{2\pi}dx  \cos^2  x= \int_0^{2\pi}dx  \sin^2  x  =\pi$.

In the limit $L\rightarrow\infty$, the index $i$ becomes
continuous $\alpha_{\nu,i}/L\rightarrow\lambda$ and the $\delta_{\nu\nu'}$
in the orthonormality condition is replaced by $\delta(\lambda-\lambda')$.
The limit of the normalisation factor is $N_k(\lambda)=d_k\lambda/(2\pi n)$,
that leads to the two-point function reported in the text (\ref{2pt}).
Note that a slightly different form of this two-point function
(satisfying Dirichlet boundary conditions) was used to investigate
the critical behaviour at an edge \cite{c-83}.

%
%
%
\end{document}